# Fast Efficient Fixed-Size Memory Pool

No Loops and No Overhead

Ben Kenwright
School of Computer Science
Newcastle University
Newcastle, United Kingdom,
*b.kenwright@ncl.ac.uk*

**Abstract--In this paper, we examine a ready-to-use, robust, and computationally fast fixed-size memory pool manager with no-loops and no-memory overhead that is highly suited towards time-critical systems such as games. The algorithm achieves this by exploiting the unused memory slots for bookkeeping in combination with a trouble-free indexing scheme. We explain how it works in amalgamation with straightforward step-by-step examples. Furthermore, we compare just how much faster the memory pool manager is when compared with a system allocator (e.g., malloc) over a range of allocations and sizes.**

## I. INTRODUCTION

A high-quality memory management system is crucial for any application that performs a large number of allocations and de-allocations. In retrospect, studies have shown that in some cases an application can spend on average 30% of its processing time within the memory manager functions [1–4] and in some cases this can be as high as 43% [5].

However, speed is only one of the features we look at for a good memory manager; in addition, we are also concerned with:

- Memory management must not use any resources (both memory or computational cost)
- Minimize fragmentation
- Complexity (ideally a straightforward and logical algorithm that can be implemented without too many problems)
- Ability to verify and identify memory problems (corruption, leaks).

Nevertheless, the majority of applications use a general memory management system, which tries to provide a best-for-all solution by catering for every possible scenario. For some systems, where speed is critical, such as games, these solutions are overkill. Instead, a simplified approach of partitioning the memory into fixed sized regions known as pools can provide enormous enhancements, such as increased speed, zero fragmentation and memory organization.

Hence, we focus on a fixed-pool solution and introduce an algorithm that has little overhead and almost no computational cost to create and destroy. In addition, it can be used in conjunction with an existing system to provide a hybrid solution with minimum difficulty. On the other hand, multiple instances of numerous fixed-sized pools can be used to produce a general overall flexible general solution to work in place of the current system memory manager.

Alternatively, in some time critical systems such as games; system allocations are reduced to a bare minimum to make the process run as fast as possible. However, for a dynamically changing system, it is necessary to allocate memory for changing resources, e.g., data assets (graphical images, sound files, scripts) which are loaded dynamically at runtime. The sizes of these resources can be determined prior to running. This then makes the fixed memory pool manager ideal. Alternatively, as mentioned a range of pools can be used for a best-fit approach to accommodate varying size data.

Naive memory pool implementations initialize all the memory pool segments when created [6][7]. This can be expensive since it is usually necessary to loop over all the uninitialized segments. Our algorithm differs by only initializing the first element and so has little computational overhead when it is created (i.e., no loops). Hence, if a memory pool is only partially used and destroyed, this wastes fewer processor cycles. Furthermore, for dynamic memory systems where partitioned memory is constantly created and destroyed this initialization cost can be important (e.g., pools being repeatedly partitioned into smaller pools at run-time).

In summary, a memory pool can make an application execute faster, give greater control, add greater flexibility, enable greater customizability, greatly enhance robustness, and prevent fragmentation. To conclude, this paper presents the implementation for a straightforward, fast, flexible, and portable fixed-size memory pool algorithm that can accomplish O(1) time complexity memory allocation and de-allocation that is ideal for high speed applications.

The fixed-size pool algorithm we present boasts the following properties:

- No loops (fast access times)
- No recursive functions
- Little initialization overhead
- Little memory footprint (few dozen bytes)
- Straightforward and trouble-free algorithm
- No-fragmentation

- Control and organization of memory

The rest of the paper is organized as follows. First, Section II discusses related work. In Section III, we outline the *contribution of the paper*, followed by Section IV, which gives a detailed explanation of how the memory pool algorithm works. Section V discusses practical implementation issues. Section VI outlines some limitations of the method. Section VIII gives some benchmark experimental results. Finally, Section IX draws conclusions and further work.

## II. RELATED WORK

The subject of memory management techniques has been highly studied over the past few decades [8–12][13]. A whole variety of techniques and algorithms are available, while some of them can be overly complex and confusing to understand. On the other hand, the technique we present here is not novel, but is a modification of an existing technique [14][6][13]; whereby loops and initialization overheads are removed; this makes the resulting algorithm extremely fast and straightforward. The method also boasts of being one of the most memory efficient implementation available since it has very little memory footprint and while giving an O(1) access time. We also give an uncomplicated implementation in C++ in the appendix.

Memory pools have been a well known choice to speed-up memory allocations/de-allocations for high-speed systems [15][16][17]. Zhao et al. [18] grouped data together from successive calls into segregated memory using memory pools to reduce pre-fetch latency. An article by Applegate [19] gave a well-defined overview of the various methods and advantages of high-performance memory in portable applications and the advantages of memory pools. Further discussion in Malakhow [20] outlines the advantages of memory pools and their applicability in high-performance multi-threaded systems.

While we present a similar single-pool allocator to Hanson [7], our algorithm is more clear-cut and makes it easier to customize for an ad-hoc implementation.

Additionally, performance considerations are discussed by Meyers [21], e.g., macros and monolithic functions, that can be applied to gain further speed-ups and gain greater reliability while incorporating good coding practices. A comparison of the computational cost of a memory management system implemented in an object orientated language (e.g., C++) is less efficient than one implemented in a functional language (e.g., C) [3][22]; however, we implemented our fixed-size memory pool in C++ because we believe it makes it more re-usable, extensible and modular.

## III. CONTRIBUTION

The contribution of this paper is to demonstrate a practical, simple, fixed-size memory pool manager that has no-loops, virtually no-memory overhead and is computationally fast. We also compare the algorithm with the standard system memory allocator (e.g., *malloc*) to give the reader a real-world computational comparison of the speed differences. The comparison emphasizes just how much faster a simple and smart algorithm can be over a more complex and general solution.

## IV. HOW IT WORKS

We explain how the fixed-size memory pool works by defining what information we have and what information we need to calculate (to help make the details more understandable, see Figure 1 and Figure 2 for illustrations).

When the pool is created, we obtain a continuous section of memory that we know the start and end address of. This continuous range of memory is subdivided into equally sized memory blocks. Each memory blocks address can be identified at this point from the start address, block-size, and the number of blocks.

This leaves the dynamic bookkeeping part of the algorithm. The algorithm must keep track of which blocks are used and un-used as they are allocated and de-allocated.

We begin by identifying each memory block using a four-byte index number. This index number can be used to identify any memory location by multiplying it by the block size and adding it to the start memory address. Hence, we have 0 to $n$-1 blocks; where $n$ is the number of blocks).

The bookkeeping algorithm works by keeping a list of the *unused* blocks. We only need to know which blocks are being unused to find the used blocks. This list of unused blocks is modified as blocks are allocated and de-allocated.

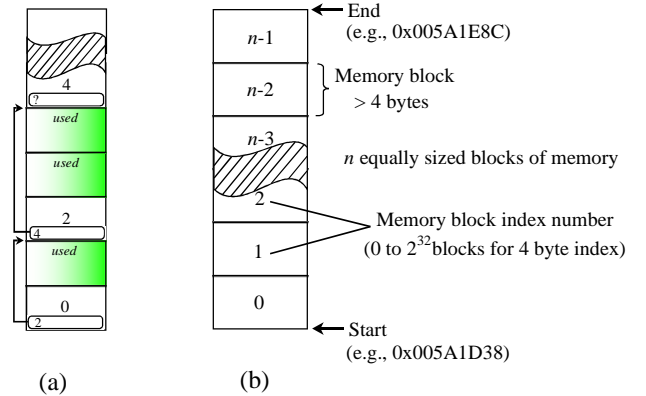

Figure 1. (a) Illustrate how the unused memory is linked together (the unused memory blocks store index information to identify the free space). (b) Example of how memory is subdivided into a number of n blocks.

However, we avoid the cost of initializing and link together all the unused blocks. We alternatively initialize a variable to inform us of how many of the $n$ blocks have been appended to the unused list. Whereby, at each allocation unused blocks are appended to the list and the number of initialized blocks variable is updated (see Figure 1).

The list uses no additional memory. Since the memory blocks that are being kept track of are not being used, we can store information inside them. Each unused block stores the index of the next unused block. The pool keeps track of the head of the unused linked chain of blocks.

For this bookkeeping algorithm to work a minimum size constraint must be imposed on the memory blocks. The individual memory blocks must be greater than four-bytes. This is because each unused memory block will hold the index of the next unused memory block to form a linked list all the unused blocks.

Therefore, each unused block holds the index to the next unused block and so on. Our pool stores the index to the head of the first unused block. For each allocation an unused block is removed from the list and returned to the user. We keep track of the head of the unused list of blocks and is updated after each allocation. Alternatively, when a block de-allocated we can calculate its index from its memory address then append it to the list of unused blocks.

We only add new unused blocks to the list during allocation. We keep track of how many blocks have been added to the list and stop appending new blocks when we have reached the upper limit. This avoids any loops and initialization costs since we only initialize blocks as we need them. In summary, as we allocate blocks, further unused blocks are initialized and appended to the list as needed.

Figure 1 is used to help further illustrate the working mechanism of the algorithm; in addition, Listing 1 gives the pseudo-code.

*A. Step-by-Step Example*

To follow the fixed-pool method through, we use a simple step-by-step example shown in Figure 2 to see the algorithm in action.

We create a fixed pool with four-blocks. We show how unused blocks and member variables change during the process of creation, allocation and de-allocation sequentially from the start (identifying uninitialized and unknown memory with question marks – the three variables in Figure 2 represent the necessary variables used by the pool for bookkeeping).

*B. Verification*

Writing a custom memory pool allocator can be both difficult and error prone. While the fixed size memory pool algorithm is relatively straightforward and trouble-free to implement, it is advised that additional verification and sanity checks be incorporated to ensure a robust and reliable implementation.

These sanity and safety checks can come at the cost of extra memory usage and increased computational cost. For example, running experimental simulations of system allocations within the debugger would increase allocation times by up to 100 times (see Figure 3 and Figure 4, which show the different allocation times of running within and outside the debugger).

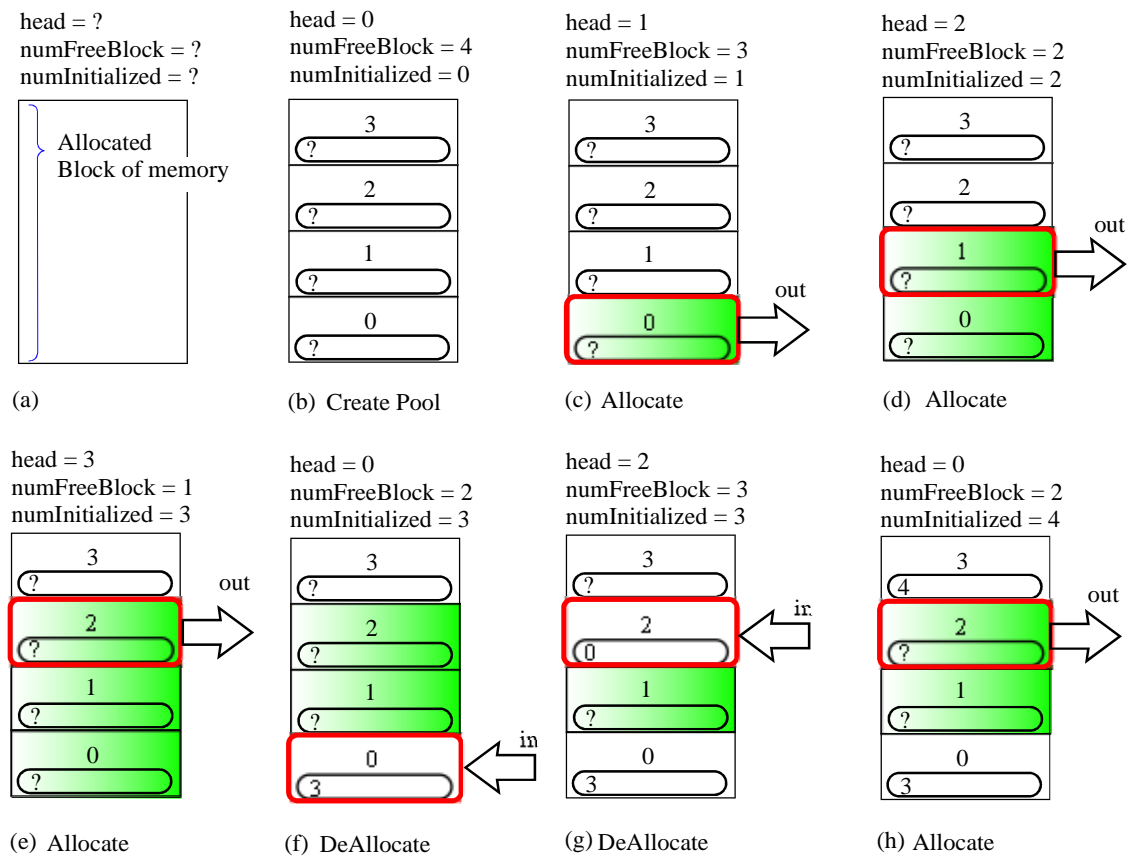

Figure 2. Step-by-step example of the memory pools internal workings for a simple 4 slot segmentation - the sequence of events from (a) to (h).

The memory pool gives the maximum amount of control and can implement various custom checks. They can be enabled and disable at will, and can be less computationally expensive than the system memory checks enabling builds to run at fast speeds while gaining debug information.

For example, the de-allocated memory addresses can easily be verified, since each memory address must be within an upper and lower boundary of the continuous memory region. Furthermore, the de-allocated memory address must be the same as one of the addresses from the divided memory blocks. In addition, memory guards can be added to include boundary checks by adding a pre and post byte signature to each block. These memory guards can be checked globally (i.e., for all blocks) and locally (i.e., currently deleted block) to identify problems and provide sanity checks.

Furthermore, leaks can be found by extending and embedding the memory guards to store additional information about the allocation; for example, the line number of the allocation.

## V. IMPLEMENTATION

We implemented the code in C++. The pool was created using create/destroy functions instead of the constructor/destructor so that the pool could be dynamically resized without destroying and recreating the pool each time it needed reconfiguring.

The implementation has four essential public functions: Create, Destroy, Allocate, and De-allocate.

The fundamental source code that implements the fixed-size memory pool is given in the appendix. To keep the source code as straightforward and as easy to read as possible all the validation and sanity check code has been excluded.

*Initialize pool*
*[ Block of memory is allocated or obtained ]*
*1. Store the start address, number of blocks and the number of uninitialized unused block*
*Allocator*
*2. Check if there any free blocks*
*3. If necessary - initialize and append unused memory block to the list*
*4. Go to the head of the unused block list*
*5. Extract the block number from the head of the unused block in the list and set it as the new head*
*6. Return the address for the old block head*
*De-allocator*
*7. Check the memory address is valid*
*8. Calculate the memory addresses index id*
*9. Set its contents to the index id of the current head of unused blocks and set itself as the head*

Listing 1. Pseudo-code for pool.

Combining the fixed pool allocator with an existing memory management system in C++ by overloading the *new* and *delete* operators would give better performance with the minimum amount of disruption, since 38% of execution time can be consumed by the dynamic memory management [3]. This ad-hoc approach works by checking the memory allocation size within the new operator; if space is available inside the pool, and the size is within a specified tolerance the memory is taken from the pool, but if not, the general system allocator is called to supply the memory.

Additionally, the greatest care must be exercised to ensure that classes and structures in C++ that are allocated and de-allocated by the fixed-size pool allocator have their constructors and destructors manually called.

## VI. LIMITATIONS

The fixed-pool memory manager relies on it being assigned a continuous block of memory. This can be a serious limiting factor if the assigned block of memory is scattered around.

Furthermore, we have focused on the algorithm and not discussed hardware limitations. For example, a page fault can result in an access time being 10,000 times slower than normal. Additionally, we have not addressed the issue of using the memory pool in a multi-threaded environment. This also raises the question of how the memory manager can be managed across multiple cores and the subject of scalability.

As well, the presented memory pool implementation is limited to systems with direct access to the memory and so cannot be implemented in managed memory environments (e.g., Java and C#).

The amount of memory requested from the fixed-size pool allocator can raise two major problems. Firstly, if the requested memory is dramatically smaller than the slot-size then lots of memory will be wasted. Secondly, and worse, if the requested memory is greater than the slot-size then it is impossible to allocate memory from the pool. Nevertheless, to combat these problems and to reduce memory wastage and largely miss-sized allocations an ad-hoc solution can be used. Whereby, a general system allocator in conjunction with multiple fixed-size pools would help to reduce memory wastage while still benefiting from the pool speedups.

On the other hand, it should be pointed out, that a general memory management system could become slower and fragmented over time. Whereby, a suitable block of memory would require considerable searching overhead, in addition to, small chunks of unsuitable and unusable memory being scattered around.

## VII. RESIZING

The fixed-size memory pool holds a list of unused memory blocks. This list resides in the unused memory and is incrementally extended when a memory block is allocated. Hence, if more memory blocks are needed than are available, and further additional memory follows the end of the continuous memory pools allocation, the pool can be extended effortlessly with little cost by updating its member variables. Once the member variables have been updated to incorporate the new end memory address it will automatically extend and fill the new region of memory during block allocations.

The algorithm currently always initializes the next unused memory block during the allocation call. However, an additional check can be added to avoid initialization of further unused blocks if they are not needed. For this reason, we could identify the maximum allocated number of unused blocks. Then, optionally the large pool of memory could be resized-down without needing to destroy and re-create the pool.

## VIII. Experimental Results

The algorithm itself is simple with no loops, no recursion, and little computational cost, and produces extremely fast allocations and de-allocations. To get a ballpark idea of how much faster the memory pool manager can be over a general memory system; we allocated and de-allocated a range of memory chunks as shown in Figure 3 and Figure 4. The figures show the fixed-pool allocator to be ten times faster than the general system allocator, and a thousand times faster when running within a debug environment.

## IX. Conclusion and Further Work

We have shown a fundamental, unsophisticated, raw-and-ready memory pool algorithm that produces remarkably fast speeds with nearly no-overhead and boasts the added advantage of being straightforward to understand and easy to implement. The fixed-size memory pool provides the best solution for processes such as games, which assume that relatively few memory allocations happen, and when they do happen they are of a deterministic size and need to be extremely fast (for example, graphical assets, particles, network packets and so on).

The Keep It Short and Simple (K.I.S.S) approach was a target goal for the fixed-size memory pool since the presented algorithm is a fundamental building block for constructing, if desired, a more elaborate and flexible memory manager.

Further work is needed to investigate if the algorithm could be optimised to use less decisional logic (i.e., if statements). In addition to exploring hardware considerations (e.g., caching, paging, registers, memory alignment, threading) and how the algorithm can be enhance to accommodate platform specific speed-ups.

# APPENDIX

## A. Fixed-Size Pool Manager - C++ Code

```
class Pool_c
{ // Basic type define
  typedef unsigned int    uint;
  typedef unsigned char   uchar;

  uint   m_numOfBlocks;      // Num of blocks
  uint   m_sizeOfEachBlock;  // Size of each block
  uint   m_numFreeBlocks;    // Num of remaining blocks
  uint   m_numInitialized;   // Num of initialized blocks
  uchar* m_memStart;         // Beginning of memory pool
  uchar* m_next;             // Num of next free block
public:

  Pool_c()
  {
    m_numOfBlocks     = 0;
    m_sizeOfEachBlock = 0;
    m_numFreeBlocks   = 0;
    m_numInitialized  = 0;
    m_memStart        = NULL;
    m_next            = 0;
  }
  ~Pool_c() { DestroyPool(); }

  void CreatePool(size_t sizeOfEachBlock,
                  uint numOfBlocks)
  {
    m_numOfBlocks     = numOfBlocks;
    m_sizeOfEachBlock = sizeOfEachBlock;
    m_memStart        = new uchar[ m_sizeOfEachBlock *
                                   m_numOfBlocks ];
    m_numFreeBlocks   = numOfBlocks;
    m_next            = m_memStart;
  }

  void DestroyPool()
  {
    delete[] m_memStart;
    m_memStart = NULL;
  }

  uchar* AddrFromIndex(uint i) const
  {
    return m_memStart + ( i * m_sizeOfEachBlock );
  }

  uint IndexFromAddr(const uchar* p) const
  {
    return (((uint)(p - m_memStart)) / m_sizeOfEachBlock);
  }

  void* Allocate()
  {
    if (m_numInitialized < m_numOfBlocks )
    {
      uint* p = (uint*)AddrFromIndex( m_numInitialized );
      *p = m_numInitialized + 1;
      m_numInitialized++;
    }

    void* ret = NULL;
    if ( m_numFreeBlocks > 0 )
    {
      ret = (void*)m_next;
      --m_numFreeBlocks;
      if (m_numFreeBlocks!=0)
      {
        m_next = AddrFromIndex( *((uint*)m_next) );
      }
      else
      {
        m_next = NULL;
      }
    }
    return ret;
  }

  void DeAllocate(void* p)
  {
    if (m_next != NULL)
    {
      (*(uint*)p) = IndexFromAddr( m_next );
      m_next = (uchar*)p;
    }
    else
    {
      *((uint*)p) = m_numOfBlocks;
      m_next = (uchar*)p;
    }
    ++m_numFreeBlocks;
  }
}; // End pool class
```

Listing 2. C++ Source Code.

## B. System Information

Simulation tests were performed on a machine with the following specifications: Windows7 64-bit, 16Gb Memory, Intel i7-2600 3.4Ghz CPU. Compiled and tested with Visual Studio.

## C. Speed Comparison Graphs

Each line represents a fixed allocation size and the time taken to allocate repeatedly.

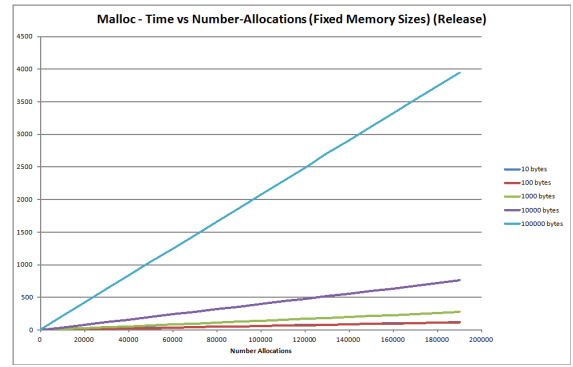

Figure 3. Release build with full optimization <u>running within the debugger</u> (Time in ms); system malloc only.

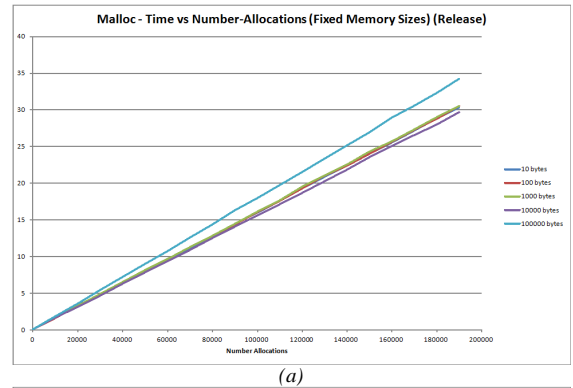

*(a)*

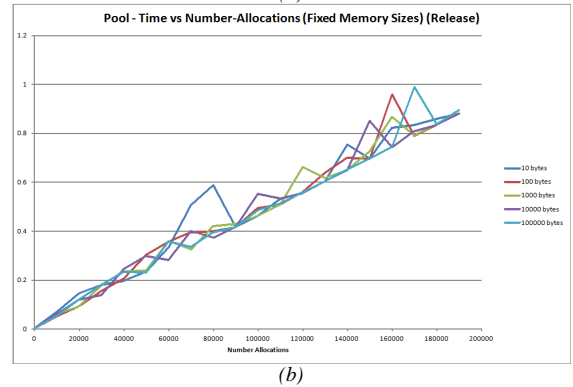

*(b)*

Figure 4. Running outside the debugger – standalone (Time in ms); (a)system malloc and, (b)custom pool.